\documentclass[11pt,twoside]{article}

\usepackage{asp2006}
\usepackage{epsf}
\usepackage{psfig}
\usepackage{lscape}

\begin{document}
\title{Isolated Galaxies and Isolated Satellite Systems}   
\author{H. B. Ann}   
\affil{Division of Science Education, Pusan National University, Busan 609-735, Korea}    
\author{Changbom Park}   
\affil{Korea Institute of Advanced Studies, Seoul 130-722, Korea}
\author{Yun-Young Choi}   
\affil{Astrophysical Research Center for the Structure and Evolution of the Cosmos, Sejong University, Seoul 143-747, Korea}

\begin{abstract} 
We search for isolated galaxies using a volume-limited sample
of galaxies with $0.02< z < 0.04742$ from SDSS DR7 supplemented by 
bright galaxies. We devise a diagnostic tool
to select isolated galaxies in different environments using the projected 
separation ($r_{\rm p}$) normalized by the virial radius of the nearest 
neighbor ($r_{\rm vir,nei}$) and the local background density. We find that
the isolation condition of $r_{\rm p}>r_{\rm vir,nei}$ and $\rho <\bar{\rho}$ 
well segregates the CIG galaxies. We confirm the morphology
conformity between the host and their satellites, which
suggests importance of hydrodynamic interaction among galaxies within their virial radii in galaxy evolution.  

\end{abstract}


\section{Introduction}   
There have been many attempts to search for isolated galaxies since the
Catalog of Isolated Galaxies (Karachentseva 1973, hereafter CIG). 
Most of the previous studies used selection criteria similar to those of 
Karachentseva (1973) who adopted diameters of galaxies and projected 
separations between a target galaxy and its neighbors. Of course, there
are several attempts to constrain isolation better by adding 
more parameter such as apparent magnitude (eg., Allam et al. 2005) and 
line of sight velocity difference (eg., Marquez \& Moles 1966) or 
by employing new criteria such as tidal strength (Varela et al. 2004).

Recent investigations of the effects of environment on the morphology of 
galaxies show that the morphology of a galaxy depends on the local background
density as well as the morphology of the nearest neighbor (Park et al. 2007, 
2008). The effects of the nearest neighbor become dominant when a target  
galaxy is located within the virial radius of the nearest neighbor.
Since the virial radius of a galaxy is assumed to be the domain of strong
tidal and hydrodynamic interactions, it can be a good parameter to select
isolated galaxies. 

The purpose of the present study is to devise a better operational criteria
for the selection of isolated galaxies using SDSS DR7 supplemented by
bright galaxies from various catalogs. We also want to demonstrate that 
the morphology of galaxies within the virial radius of a galaxy is likely
to resemble each other through the hydrodynamical interactions. 

\section{Virial Radius and Local Background Density}  
The morphology of a galaxy depends on the local background density as well 
as the morphology of the nearest galaxy (Park et al. 2007, 2008) but the 
effects of the nearest neighbor are dominated within the virial radius of the
nearest neighbor (Park et al. 2008, Park \& Choi 2009). We calculate the
virial radius of a galaxy, 
$$r_{\rm vir}=(3\gamma L/{4\pi \rho_{c}}/200)^{1/3}$$ 
where $\rho_{c}$ is 
the critical density of the universe and $\gamma$ is the mass--to--light
ratio (Park et al. 2008). 
We adopted $\gamma=2$ for early-type galaxies and $\gamma=1$ for 
late-type ones. 
The local background density is calculated as
$$ \rho = 7/{4\pi r_{p}^{2}}$$
where $r_{p}$  is projected distance to the 7th nearest galaxy
that is brighter than $M_r = -19.5$ 
with line--of--sight velocity difference less than 1000 km/s. 

\section {Isolated Galaxies}
\subsection {CIG Galaxies}
Fig. 1 shows the distribution of the primary sample galaxies
in the log(${r_{\rm p}/r_{\rm vir,nei}})$ versus 
log(${\rho/\bar{\rho}}$) 
diagram where $r_{\rm vir,nei}$ is the virial radius of 
the nearest neighbor galaxy and $\bar{\rho}$ is the mean local background
density of the primary sample. 

\begin{figure}
\centerline{\psfig{figure=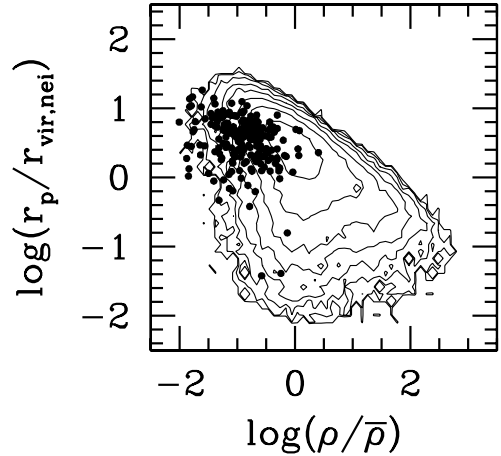,height=6cm}}
\caption{{Distribution of galaxies in the 
log(${r_{\rm p}/r_{\rm vir,nei}}$) versus log(${\rho/ \bar{\rho}}$) diagram. 
The CIG galaxies are plotted as solid circles and the distribution of
the volume-limited sample of galaxies are represented
by the iso-number contours. 
}}
\end{figure}

There is an upper limit of log(${r_{\rm p}/r_{\rm vir,nei}}$) for galaxies at a 
local background density which seems to be set by the anti-correlation 
between the neighbor distance and the local background density.
The other two boundaries of the distribution 
are not well defined but they reflect the wide ranges of separation relative
to the neighbor's size for the galaxies in low local densities and the 
wide ranges of local densities for galaxies with small relative separations.
There are several interesting features in Fig. 1, but here we note that 
the lower limit of the relative separation of galaxies in the lower right 
part of the diagram is thought to be caused
by the cannibalism prevalent in the densest regions.  

As shown in Fig 1, most of CIG galaxies are located under-dense regions
with $\rho<\bar{\rho}$.
Only a small fraction of CIG galaxies are highly isolated galaxies 
with $r_{\rm p}/r_{\rm vir,nei} > 10$. Since CIG galaxies, which are thought
to be a representative sample of isolated galaxies, are well 
segregated from others in log(${r_{\rm p}/r_{\rm vir,nei}}$)
versus log(${\rho/ \bar{\rho}}$) diagram, the virial radius when it is used 
with the local background density seems to be a good 
selection criterion for isolation of a galaxy. 

\subsection{Galaxies in Extreme Isolation}   
Galaxies in extreme isolation is of interest for many reasons. In Figure 2, we 
present a relation between the luminosity and color of galaxies 
that have no companion within 2 Mpc. We determined  morphology of galaxies
using the automated morphology classifier that is devised specifically 
for SDSS galaxies (Park \& Choi 2005).  As shown in Figure 2, late-type 
galaxies (small solid circle) show a smooth distribution while early types 
(triangles) display a bimodal distribution centered at $u-r \approx 1.5$
and $u-r \approx 2.6$, respectively for the blue and red groups.
Early-type galaxies in the former group are mostly blue 
elliptical galaxies of which a large 
fraction is dwarf elliptical galaxies. Galaxies in the 
latter group consist mostly of normal elliptical galaxies and their 
distribution resembles the red sequence observed in the local universe.
 
\begin{figure}
\centerline{\psfig{figure=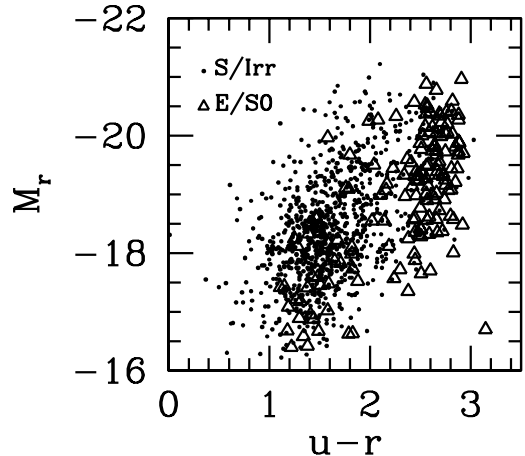,height=6cm}}
\caption{Luminosity and color of extremely isolated galaxies.
}

\end{figure}

\subsection{Isolated Satellite Systems}
A galactic satellite system consists of a central host galaxy and its
satellite galaxies which are fainter than their host galaxy. We follow the
method of Ann et al. (2008) to find out isolated satellite systems. 
We found about three times more satellite systems than Ann et al. (2008) due
to an enlarged sample size. 
We confirmed the morphology conformity between host galaxy and its satellites
found by Ann et al. (2008) in the present enlarged sample.
As suggested by Ann et al. (2008), the morphology conformity in galactic
satellite systems is believed to be caused mainly by the hydrodynamical 
interaction between hosts and satellites,

\section{Discussion and Conclusions}
We devised a diagnostic tool for
selection of isolated galaxies using relative separation ($r_{\rm p}/r_{\rm vir, nei}$)
and the local background density.
We showed that CIG is a really good representative sample of isolated 
galaxies since almost all the CIG galaxies in common with SDSS DR7 are located
in under-dense regions, with the projected distance to the nearest 
neighbor ($r_{\rm p}$) greater than the nearest neighbor's virial 
radius ($r_{\rm vir, nei}$).
The success of Karachentseva's criteria seems to be
due to the proper consideration of the projected separation relative to the 
size of the companion. 

We derived a subset of isolated galaxies that have no companion galaxy
within 2 Mpc, most of which is thought to be located in the extremely 
under-dense regions like voids. There is a bimodal distribution of early
type galaxies in this region in the $M_{r}$ versus $u-r$ diagram. The early 
type galaxies with red colors follows the red sequence and are thought 
to be normal elliptical galaxies while the early type galaxies with blue
colors are dominated by dwarf ellipticals. 

We confirm the morphology conformity in galactic satellite systems found
by Ann et al. (2008). Since morphology conformity is more prevalent inside
the virial radius of host galaxy, it seems to be mainly driven by 
hydrodynamical interactions between the hosts and satellites 
that lead to the transformation of their morphology.
 
\acknowledgements 
This work was supported by the research grant of KOSEF through ARCSEC.


\end{document}